\documentclass[prd,twocolumn,showpacs,reprint,preprintnumbers,nofootinbib,amsmath,amssymb]{revtex4-2}

\RequirePackage[colorlinks=true
,urlcolor=blue
,anchorcolor=blue
,citecolor=blue
,filecolor=blue
,linkcolor=blue
,menucolor=blue
,linktocpage=true
,pdfproducer=medialab
,pdfa=true
]{hyperref}

\usepackage{amsmath,amssymb,amsthm,amsfonts}
\usepackage{bbold}
\usepackage{graphicx,tabularx}
\usepackage{color}
\usepackage{multirow}  
\usepackage{wasysym}  
\usepackage{comment}
\usepackage{slashed}
\usepackage{enumitem}
\usepackage{cleveref}
\usepackage{rotating}
\usepackage[english]{babel}
\usepackage[justification=centerlast]{caption}
\usepackage{subcaption}
\usepackage{float}

\usepackage{breqn} 
\makeatletter 
\let\cref@old@eq@setnumber\eq@setnumber 
\def\eq@setnumber{%
\cref@old@eq@setnumber%
\cref@constructprefix{equation}{\cref@result}%
\protected@xdef\cref@currentlabel{%
[equation][\arabic{equation}][\cref@result]\p@equation\theequation}} 
\makeatother

\usepackage{cases}

\crefname{section}{Sec.}{Secs.}
\crefname{figure}{Fig.}{Figs.}
\crefname{equation}{Eq.}{Eqs.}
\crefname{appendix}{Appendix}{Appendices}
\setlist[description]{leftmargin=0.4cm}
\setlist[itemize]{leftmargin=0.4cm}

\newcommand{\be}{\begin{equation}\begin{aligned}}
\newcommand{\ee}{\end{aligned}\end{equation}}

\newcommand{\beq}{\begin{equation}}
\newcommand{\eeq}{\end{equation}}
\newcommand{\beqa}{\begin{eqnarray}}
\newcommand{\eeqa}{\end{eqnarray}}

\renewcommand{\eqref}[1]{Eq.~(\ref{#1})}

\newcommand{\eg}{{\em e.g.}}
\newcommand{\ie}{{\em i.e.}}

\RequirePackage[normalem]{ulem}

\DeclareUnicodeCharacter{2212}{\textendash}

\makeatletter
\def\l@subsubsection#1#2{}
\makeatother

\usepackage{multirow}
\usepackage{makecell}

\usepackage{fontawesome}

\usepackage[T1]{fontenc}

\begin{document}

\preprint{CTPU-PTC-24-37}

\title{Covariant quantum field theory of tachyons is unphysical}

\author{Krzysztof Jod\l{}owski}
\email{k.jodlowski@ibs.re.kr}
\affiliation{Particle Theory and Cosmology Group\char`,{} Center for Theoretical Physics of the Universe\char`,{} Institute for Basic Science (IBS)\char`,{} Daejeon\char`,{} 34126\char`,{} Korea}

\begin{abstract}
    Tachyons have fascinated generations of physicists due to their peculiar behavior, but they did not solve any real physical problem. This may have changed with the recent works of Dragan et al., who have shown that superluminal observers may be related to the foundations of quantum mechanics (QM) because they require introducing non-determinism and wave-like behavior at the fundamental level. Unfortunately, we show that the tachyon quantum field theory proposed as a part of this program is not quantum since the tachyon field commutes at all points and the canonical commutations relations of a quantum scalar field are not satisfied. We also discuss other authors' theories, \eg, we show that the Dhar-Sudarshan's formula for the Feynman propagator violates unitarity, and we apply the LSZ formalism to tachyons for the first time, where we find that one cannot prove the LSZ asymptotic condition just by replacing plane waves with wave packets.
\end{abstract}

\renewcommand{\baselinestretch}{0.85}\normalsize
\maketitle
\renewcommand{\baselinestretch}{1.0}\normalsize

\section{Introduction}
\label{sec:introduction}
Many works have investigated tachyons, superluminal particles characterized by a negative mass squared parameter, due to their peculiar behavior, \eg, causality paradoxes or nondeterministic behavior; see the discussion and references given in Ref.~\cite{Paczos:2023mof}. The challenges facing consistent description of tachyons (\eg, group theory precludes the democratic treatment of subluminal and superluminal observers~\cite{PhysRevD.27.1740}) have recently been overcome~\cite{Dragan:2019grn} at the cost of interpreting the 1+3 spacetime of a subluminal observer as a 3+1 spacetime of a superluminal observer (in particular, subluminal and superluminal observers are distinguishable). However, these two families of observers are still postulated to be equivalent in the sense of the ``Quantum principle of relativity'' (QPR).

An interesting observation of~\cite{Dragan:2019grn} is that superluminal observers actually require the introduction of nondeterministic physics similar to QM, \eg, a wave-like - instead of point-like - description of Nature using complex numbers. Moreover, it was shown that proper formulation of superluminal reference frames must invoke field theory, and both classical~\cite{Dragan:2022txt} and quantum~\cite{Paczos:2023mof} field theory (QFT) of tachyons have been developed. Since QFT is the most fundamental description of physical interactions, the \textit{ultimate} test of such a program is to build a consistent covariant QFT of tachyons, which Ref.~\cite{Paczos:2023mof} has claimed to accomplish.

We note that several works~\cite{DelSanto:2022oku,Grudka:2021fdq,Horodecki:2023hvf,Grudka:2023mrp} have already raised significant questions about the results presented in~\cite{Dragan:2019grn}.
While these critiques have received responses~\cite{Dragan_2022,Dragan_2023,Dragan:2023kpa}, 
it seems that it has already been established
that there are fundamental issues with QPR. 
For example, there is an agreement that the full-fledged QM has not been recovered yet~\cite{Horodecki:2023hvf,Dragan:2023kpa} and there is insufficient explanation for different types of quantum randomness~\cite{Horodecki:2023hvf}. 
For completeness, we also note that the claim that invoking superluminal observers can explain quantum superposition has been challenged in~\cite{Grudka:2023mrp}.
Therefore, with respect to QPR, our work can be viewed as complementary to~\cite{DelSanto:2022oku,Grudka:2021fdq,Horodecki:2023hvf,Grudka:2023mrp}, where we aim to comprehensively analyze both the relativistic and QM aspects of tachyons within the most successful physical framework to date - QFT. 
On the other hand, this also means that our arguments do not apply to the results of~\cite{Dragan:2019grn,Dragan:2022txt}.

By analyzing the commutation relations in the twin space, we will show that the theory constructed in Ref.~\cite{Paczos:2023mof} is in fact not a quantum theory. Moreover, since it could be interpreted that this paper has adapted a Feynman propagator (FP) from an earlier result due to Dhar and Sudarshan (DS)~\cite{Dhar:1968hkz}, we explicitly show that the DS FP violates unitarity. Our work also has an impact on~\cite{Paczos:2024mfx}, since it assumed the results of Ref.~\cite{Paczos:2023mof}, some of which we question.

\section{The twin space}
\label{sec:twin_space}
Since the properties of the twin space introduced in Ref.~\cite{Paczos:2023mof} play a key role in the tachyon quantization performed in that paper, and many of the twin space properties and technical details were only stated implicitly, we start by elucidating its properties.
The twin space is ${\cal F} \otimes {\cal F}^\star$, where ${\cal F}$ is the standard Fock space for a quantum scalar field \cite{Weinberg_1995} and ${\cal F}^\star$ is its dual (in the sense of vector space, \ie, it is a space of linear functionals instead of vectors \cite{axler15}). 
We note that Ref.~\cite{Paczos:2023mof} has not explicitly shown that their construction is mathematically well-defined\footnote{We thank the anonymous Referee for emphasizing this point.} - in particular, there is no explicit formula for $\phi^\star$ in Ref.~\cite{Paczos:2023mof}. 
On the other hand, the creation $\hat{a}^{\star}_{\vec{k}}$ and annihilation $\hat{a}^{\star\dagger}_{\vec{k}}$ operators in the dual space have been properly defined, which is as follows~\cite{Paczos:2023mof}: 
\be
\hat{a}^\star_{\vec{k}}\, \langle 0|\, \equiv \langle \vec{k} |\,, \,\,\,\, \hat{a}^{\star\dagger}_{\vec{k}}\, \langle 0|\,\equiv 0\,,
\label{eq:norm_dual}
\ee
where $\langle \vec{k} |$ is momentum $\vec{k}$ eigenstate and $\langle 0|$ is the ground state in the dual space.

Let us start by noting that the twin space can be identified with the Fock space of the direct sum of two single-particle Hilbert spaces using the exponential law for Fock spaces, see Section 5.6 in Ref.~\cite{2006LNP...695...63D}. Therefore, it is a bona fide Hilbert space that by itself does not need further mathematical justification.

The main object of study in Ref.~\cite{Paczos:2023mof} is the tachyon quantum field 
$\hat{\Phi}: {\cal F} \otimes {\cal F}^\star \to {\cal F} \otimes {\cal F}^\star$, which takes the following form:
\be
    \hat{\Phi}(x) &= \frac12 \left(\hat{\phi}(x) \otimes \hat{\mathbb{1}} + \hat{\mathbb{1}} \otimes \hat{\phi}^{\star}(x)\right)\,,
\label{eq:app_1}
\ee
where
\be
\hat{\phi}(x) = \int \frac{d^3\vec{k}}{(2\pi)^3}\, \theta(|\vec{k}|-m) \, \left( u_{\vec{k}}(x)\, \hat{a}_{\vec{k}} + u^*_{\vec{k}}(x)\, \hat{a}^\dagger_{\vec{k}} \right)\,,
\label{eq:app_phi}
\ee
and
\begin{equation}
    u_{\vec{k}}(x)=\frac{e^{-i\omega_{\vec{k}} t+i\vec{k}\cdot\vec{x}}}{(2\pi)^2 \,2\omega_{\vec{k}}} \,, \,\,\,\, \mathrm{ } \,\,\,\, [\hat{a}_{\vec{p}},\hat{a}^\dagger_{\vec{q}}]=(2\pi)^3 \, 2\omega_{\vec{p}} \, \delta^3(\vec{p}-\vec{q})\,,
\label{eq:CCR}
\end{equation}
where the RHS of \cref{eq:CCR} are the canonical commutation relations (CCRs) of $a$, $a^\dagger$ in the ordinary Fock space $\cal F$.

To obtain the CCRs in the dual space, one can use the identity 
\be
(A \circ B)^\star = B^\star \circ A^\star \,,
\label{eq:dual_id}
\ee
where $A$, $B$ act on the space ${\cal F}$ and $A^\star$, $B^\star$ act on ${\cal F}^\star$ \cite{axler15}.
Applying the star operator to the RHS of \cref{eq:CCR}, Ref.~\cite{Paczos:2023mof} obtained
\begin{equation}
    [\hat{a}^{\dagger\star}_{\vec{q}},\hat{a}^\star_{\vec{p}}]=(2\pi)^3 \, 2\omega_{\vec{p}} \, \delta^3(\vec{p}-\vec{q})\, .
\label{eq:app_8}
\end{equation}
Note that this is consistent with \cref{eq:norm_dual}, since the same relation can be derived by demanding that the momentum eigenstates in the dual space are relativistically normalized, $\langle \vec{l}\,|\vec{k}\rangle_{F^\star}=(2\pi)^3\, 2\omega_k \,\delta(\vec{k}-\vec{l})$.
Therefore, the ``starred'' creation/annihilation operators indeed satisfy the identity given by \cref{eq:dual_id}.

Ref.~\cite{Paczos:2023mof} does not include an explicit formula for $\hat{\phi}^{\star}(x)$, which has lead to some confusion.
Let us define two inequivalent forms of $\hat{\phi}^{\star}(x)$, which we denote as $\hat{\phi}_1^{\star}(x)$ and $\hat{\phi}_2^{\star}(x)$, respectively:
\be
    &\hat{\phi}_1^{\star}(x) \equiv \int \frac{d^3\vec{k}}{(2\pi)^3}\, \theta(|\vec{k}|-m) \, \left( u_{\vec{k}}(x)\, \hat{a}^\star_{\vec{k}} + u^*_{\vec{k}}(x)\, \hat{a}^{\star\dagger}_{\vec{k}} \right)\,,\\
    &\hat{\phi}_2^{\star}(x) \equiv \int \frac{d^3\vec{k}}{(2\pi)^3}\, \theta(|\vec{k}|-m) \, \left( u^*_{\vec{k}}(x)\, \hat{a}^\star_{\vec{k}} + u_{\vec{k}}(x)\, \hat{a}^{\star\dagger}_{\vec{k}} \right)\,,\\
\label{eq:phi_12}
\ee
where the $*$ symbol denotes complex conjugation and $\theta(\cdot)$ is the Heaviside step function. 
Note that $\hat{\phi}_2^{\star}(x)$ acts like a Hilbert adjoint~\cite{Muscat} of $\hat{\phi}(x)$, that is it acts like $\hat{\phi}^\dagger(x)$.
On the other hand, $\hat{\phi}_1^{\star}(x)$ lacks complex conjugation on the complex numbers $u_{\vec{k}}(x)$ inside the $\hat{\phi}(x)$ operator, hence it acts more like a transpose of $\hat{\phi}(x)$, $\hat{\phi}^T(x)$.
Actually, both the adjoint $\dagger$ and the transpose $T$ satisfy the identity satisfied by the $\star$ in \cref{eq:dual_id}.
We will show that adapting $\hat{\phi}^\star=\hat{\phi}_1^\star$ leads to the claims given in Section II of Ref.~\cite{Paczos:2023mof}, \textit{except} that the tachyon field $\hat{\Phi}$ satisfies the CCR of a quantum scalar - this means that while $\hat{\Phi}$ is covariant, it is not a quantum field.\footnote{Later, we show that if $\hat{\phi}^\star=\hat{\phi}_2^\star$, then the field $\hat{\Phi}$ violates microcausality and its commutator is not LI, similarly to~\cite{Dhar:1968hkz}.}
By two different methods, we will show that $\hat{\Phi}$'s commutator vanishes for any points $x$, $y$ in Minkowski space,\footnote{Actually,~\cite{Jue:1973mr} has already proposed a tachyon QFT with such a property, which was interpreted that tachyons are only virtual states. However, the vanishing of the commutator, which was adapted from~\cite{Murphy:1972ii}, relies on assuming the full LI measure $d^4k$, instead of $d^4k \, \theta(k^0)$. Actually, the proof given in~\cite{Murphy:1972ii} did not use the assumption on the sign of $k^2$, hence it actually proved a statement about the LI measure and not about tachyons. Note that this remark is not relevant to Ref.~\cite{Paczos:2023mof}, which used $d^4k \, \theta(k^0)$.}
\begin{equation}
[\hat{\Phi}(x), \hat{\Phi}(y)] = 0\,,
\label{eq:Phi_Phi}
\end{equation}
and it does not satisfy the CCR for a quantum field~\cite{Weinberg_1995,2006LNP...695...63D}, \ie, 
\begin{equation}
    [\hat{\Phi}(0,\vec{x}), \partial_t \hat{\Phi}(0,\vec{y})|_{t=0}] = 0 \neq i \delta(\vec{x}-\vec{y}).
\label{eq:Phi_Pi}
\end{equation}

Before showing that, let us discuss the covariance of $\hat{\Phi}$ first, since Ref.~\cite{Paczos:2023mof} sketched it. 
We need to show that $\hat{\Phi}$ satisfies the Lorentz transformation law for a scalar field \cite{Weinberg_1995}
\be
    \hat{\Phi}(\Lambda^{-1} x) = U(\Lambda)^{-1}\, \hat{\Phi}(x)\, U(\Lambda),
\label{eq:app_2}
\ee    
where $\Lambda$ is any transformation belonging to the connected component of the Lorentz group ($\mathrm{det}\, \Lambda=1$ and $\Lambda^0_{\,0}>0$), and $U$ is its unitary representation on the twin space.
The extension of the usual Hilbert space to twin space, on which the tachyon quantum field acts on, is needed, because neither of the fields $\hat{\phi}$ and $\hat{\phi}^{\star}$, which act on ${\cal F}$ and ${\cal F}^\star$, respectively, are covariant on their own \cite{Paczos:2023mof}.
Indeed, Lorentz boost for space-like four vectors transforms an annihilation operator into creation operator, which does not preserve the canonical commutation relations - see the discussion around Eq. 6 in~\cite{Paczos:2023mof}.

We first show that the tachyon field $\hat{\phi}$ in the $\cal F$ space  has the following form  in the momentum representation:
\be
    \hat{\phi}(k) &= (2\pi)\, \theta(|\vec{k}|-m) \, e^{-ikx}  \, \delta(k^2+m^2) \\
    &\times\, \left(\theta(k^0)\hat{a}_{\vec{k}} + \theta(-k^0)\hat{a}^\dagger_{-\vec{k}}\right).
\label{eq:app_3}
\ee    
We use \cref{eq:app_phi}, the identity $\delta(k^2+m^2)=\frac{\delta(k^0-\omega_{\vec{k}})+\delta(k^0+\omega_{\vec{k}})}{|2k^0|}$, where $\omega_{\vec{k}}=\sqrt{|\vec{k}|^2-m^2}$, and the invariance of the measure under $\vec{k}\to-\vec{k}$:
\begin{widetext}
\be
    \hat{\phi}(x) &= \int \frac{d^3\vec{k}}{(2\pi)^3}\, \theta(|\vec{k}|-m) \, \left( u_{\vec{k}}(x)\, \hat{a}_{\vec{k}} + u^*_{\vec{k}}(x)\, \hat{a}^\dagger_{\vec{k}} \right) = \int \frac{d^3\vec{k}}{(2\pi)^3}\, \theta(|\vec{k}|-m) \, \left(\frac{e^{-i\omega_{\vec{k}} x^0  +i\vec{k}\cdot \vec{x}}}{2 \omega_{\vec{k}}} \hat{a}_{\vec{k}} + \frac{e^{i\omega_{\vec{k}} x^0  +i\vec{k}\cdot \vec{x}}}{2 \omega_{\vec{k}}}\hat{a}^\dagger_{-\vec{k}}\right) \\
    &= \int \frac{d^3\vec{k}}{(2\pi)^3} \int dk^0 \, \theta(|\vec{k}|-m) \, e^{-ikx}  \, \frac{\delta(k^0-\omega_{\vec{k}})+\delta(k^0+\omega_{\vec{k}})}{|2k^0|} \, \left(\theta(k^0)\hat{a}_{\vec{k}} + \theta(-k^0)\hat{a}^\dagger_{-\vec{k}}\right) \\
    &= \int \frac{d^4k}{(2\pi)^4} (2\pi)\, \theta(|\vec{k}|-m) \, e^{-ikx}  \, \delta(k^2+m^2)\, \left(\theta(k^0)\hat{a}_{\vec{k}} + \theta(-k^0)\hat{a}^\dagger_{-\vec{k}}\right)\, .
\label{eq:app_4}
\ee
\end{widetext}

Next, we need to show that the condition $|\vec{k}|\,>m$ is Lorentz invariant (LI) for an on-shell tachyon moving with velocity $\vec{v}=\vec{k}/E=\vec{k}/\omega_{\vec{k}}$. We use Eq. 22 from~\cite{Dragan:2022txt} to obtain the magnitude of the momentum $\vec{k'}$ after applying a subluminal boost with velocity $\vec{u}$. We need to show that
\be
    |\vec{k'}|=\frac{|\vec{k}-E \vec{u}|}{\sqrt{1-u^2}} \geq \frac{|\vec{k}|-|E| u}{\sqrt{1-u^2}} =  \frac{m(v-u)}{\sqrt{1-u^2}\sqrt{v^2-1}} \stackrel{?}{\geq} m \,.
\label{eq:LI_onshell_Heavyside}
\ee
Since $v=\left|\vec{v}\right|>1>u=\left|\vec{u}\right|$, we can square both sides and show that our condition is equivalent to the true statement $(1-uv)^2 \geq 0$. We used the reverse triangle inequality for the length of 3-momentum, and the on-shell condition by using $|\vec{k}/E|=v$. 
On the other hand, an off-shell tachyon with $E=|\vec{k}/u|$ has $|\vec{k'}|=0$, and therefore $\theta(|\vec{k}|-m)$ is not LI in general.

Demanding that condition given by \cref{eq:app_2} holds for $\hat{\phi}(x)$, gives:
\be
    &\theta(|\vec{l}|-m) \left(\theta(l^{0})\,\hat{a}_{\vec{l}} + \theta(-l^{0})\,\hat{a}^\dagger_{-\vec{l}} \right)  = \\
    &\theta(|\vec{k}|-m) \left( U(\Lambda) \, \hat{a}_{\vec{k}}\, U(\Lambda)^{-1} + U(\Lambda) \,\hat{a}^\dagger_{-\vec{k}}\, U(\Lambda)^{-1} \right) \, ,
\label{eq:app_5}
\ee
where $l=\Lambda k = (l^0,\vec{l}\,)$. Since we consider on-shell tachyon field, \cref{eq:LI_onshell_Heavyside} holds, so we can drop the terms $\theta(|\vec{l}|-m)$ and $\theta(|\vec{k}|-m)$ in \cref{eq:app_5}.
For subluminal scalar, elements of the connected part of the Lorentz group preserve the sign of the energy, \ie, if $k^{0}>0$, then $l^{0}>0$, because their energy four-vector is time-like (by the same token, they also preserve the time flow direction). 
For such transformations \cref{eq:app_5} implies that
\be
    \hat{a}_{\vec{l}} = U(\Lambda) \, \hat{a}_{\vec{k}}\, U(\Lambda)^{-1}, \,\,\, \mathrm{and} \,\,\,\, \hat{a}^\dagger_{\vec{l}} = U(\Lambda) \, \hat{a}^\dagger_{\vec{k}}\, U(\Lambda)^{-1}.
\label{eq:app_6}
\ee
This shows covariance of subluminal scalar quantum field since the commutation relations, given by \cref{eq:CCR}, are clearly preserved under action of $\Lambda$.
On the other hand, tachyons energy four-vector is space-like, hence boosts that change the sign of the energy, $l^0<0$ for $k^0>0$ exist.
In this case, \cref{eq:app_5} implies that
\be
    \hat{a}_{\vec{l}} = U(\Lambda) \, \hat{a}^\dagger_{-\vec{k}}\, U(\Lambda)^{-1} \,\,\, \mathrm{and} \,\,\,\, \hat{a}^\dagger_{\vec{l}} = U(\Lambda) \, \hat{a}_{-\vec{k}}\, U(\Lambda)^{-1} \,,
\label{eq:app_7}
\ee
instead.
This shows that superluminal scalar quantum field that satisfy \cref{eq:app_2} do not preserve the canonical commutation relations.
Under the Lorentz boost $\Lambda$ that changes the sign of the energy, the following holds:
\be
    &[\hat{a}_{\vec{p}},\hat{a}^\dagger_{\vec{q}}] = (2\pi)^3 \, 2\omega_{\vec{p}} \, \delta^3(\vec{p}-\vec{q}) \\
    &\stackrel{\Lambda}{\to} \,\,\, [\hat{a}^\dagger_{-\vec{\Lambda p}},\hat{a}_{-\vec{\Lambda q}}] = -(2\pi)^3 \, 2\omega_{\vec{\Lambda p}} \, \delta^3(\vec{\Lambda p} - \vec{\Lambda q}) \,.
\label{eq:app_tmp}
\ee

The solution proposed in Ref.~\cite{Paczos:2023mof} is to replace $\hat{\phi}(x)$ with $\hat{\Phi}(x)$, which corresponds to replacing $\hat{a}_{\vec{k}} \to \hat{c}_{\vec{k}} \equiv \hat{a}_{\vec{k}} \otimes \hat{\mathbb{1}} + \hat{\mathbb{1}} \otimes \hat{a}^\star_{\vec{k}}$.
This means that $\hat{\Phi}(x)=\hat{\Phi}_1(x)\equiv \frac12 \left(\hat{\phi}(x) \otimes \hat{\mathbb{1}} + \hat{\mathbb{1}} \otimes \hat{\phi}_1^{\star}(x)\right)$.
Direct calculation gives 
\begin{equation}
    [\hat{c}_{\vec{k}}, \hat{c}^\dagger_{\vec{l}}] = [\hat{c}_{\vec{k}}, \hat{c}_{\vec{l}}\,] = 0\,,
\label{eq:ccdagger}
\end{equation}
for any $\vec{k}$ and $\vec{l}$.
In fact, \cref{eq:ccdagger} implies \cref{eq:Phi_Phi} and \cref{eq:Phi_Pi} by direct computation.\footnote{Taking the derivative of \cref{eq:Phi_Phi} with respect to $y^0=0$, we obtain \cref{eq:Phi_Pi}, so any theory with everywhere vanishing commutators will not satisfy CCR.}

Actually, no computation is needed (except to cross-check), since we can invoke \cref{eq:dual_id}, the fact that the commutator\footnote{The same holds for two-point time-ordered correlation function.} for free field $\hat{\phi}$ is a c-number, and the fact that $\star$ operator (note we assumed that $\hat{\phi}^\star=\hat{\phi}_1^\star$) is an identity when acting on complex numbers:
\be
    &[\hat{\Phi}(x), \hat{\Phi}(y)] =  \frac12 \left( [\hat{\phi}(x), \hat{\phi}(y)] + [\hat{\phi}^\star(x), \hat{\phi}^\star(y)] \right) =\\
    &=\frac12 \left( [\hat{\phi}(x), \hat{\phi}(y)] - [\hat{\phi}(x), \hat{\phi}(y)]^\star \right) =\frac12 \left( C - C^\star\right) = \\
    &=\frac12 \left( C - C\right)  = 0 \,,
\label{eq:Phi_Phi_1}
\ee
where $C=[\hat{\phi}(x), \hat{\phi}(y)]$ is a c-number.
Moreover, since 
\begin{align}
&[\hat{\phi}(0,\vec{x}), \partial_t \hat{\phi}(0,\vec{y})|_{t=0}] = i \delta(\vec{x}-\vec{y}) \nonumber \\
& + i \, \frac{\sin{(m|\vec{x}-\vec{y}|)}-m|\vec{x}-\vec{y}|\,\cos{(m|\vec{x}-\vec{y}|)}}{2\pi^2 |\vec{x}-\vec{y}|^3} \,,
\end{align}
which is also a c-number, \cref{eq:Phi_Pi} holds by the same token.

In fact, the above results are not surprising, since the tachyon field can be written explicitly in the following form:
\be
\hat{\Phi}(x) = \int \frac{d^3\vec{k}}{(2\pi)^3}\, \theta(|\vec{k}|-m) \, \left( u_{\vec{k}}(x)\, \hat{c}_{\vec{k}} + u^*_{\vec{k}}(x)\, \hat{c}^\dagger_{\vec{k}} \right) = \hat{\Phi}^\dagger(x)\,,
\ee
where $[\hat{c}_{\vec{k}}, \hat{c}^\dagger_{\vec{l}}] = [\hat{c}_{\vec{k}}, \hat{c}_{\vec{l}}\,] = 0$.
Therefore, the covariance of $\Phi=\Phi_1$ takes place because everything commutes, which is preserved by boosts. 

If $\hat{\Phi}(x)=\hat{\Phi}_2(x)$ is adapted instead, then 
\be
    &[\hat{\Phi}(x), \hat{\Phi}(y)] = \frac12 \left( C - C^\star\right) =\frac12 \left( C - C^*\right)  =i\, \mathrm{Im}(C)=C \\
    &=\int_{|\vec{k}|>m} \frac{d^4 k}{(2\pi)^4} \,\delta(k^2+m^2) \, e^{-i k(x-y)} \left(\theta(k^0) - \theta(-k^0)\right) \,,
\label{eq:Phi_Phi_aaaa}
\ee
which has the same form as for the $m^2>0$ scalar, except for the $|\vec{k}|\,>m$ condition - which is LI on-shell by \cref{eq:LI_onshell_Heavyside} - and the sign in the mass squared term under the Dirac delta. The second property means that for tachyons the sign of the $k^0$, the $\left(\theta(k^0) - \theta(-k^0)\right)$ term, is not LI.
Therefore, \cref{eq:Phi_Phi_aaaa} is not LI, it vanishes only for $x^0=y^0$,\footnote{For tachyons, this can also be interpreted as an infinitely fast moving tachyon, where there is no dynamics because everything happens at an instance. Moreover, motion at infinite speed is not a LI property.} and it does not satisfy microcausality condition, which may cause problems with causality and prohibits constructing tachyon scattering theory since the time-ordered correlation functions (TOCF) are not well-defined (time-ordering is not LI). 

Let us also mention that the form of the FP within the theory of Ref.~\cite{Paczos:2023mof} is unclear. It was claimed that ``We also applied our framework to account for interactions with other fields''. 
However, the Eq. 21 given therein clearly cannot be used, since - as is noted in Ref.~\cite{Paczos:2023mof} - the RHS of Eq. 21 is not LI; also see the discussion under \cref{eq:LI_onshell_Heavyside}.
In fact, Ref.~\cite{Paczos:2023mof} mentioned that DS proposed to restore the LI by also including the $|\vec{k}|<m$ modes~\cite{Dhar:1968hkz}. 
However, it is not stated clearly, which form of the FP should be used in perturbative calculations of the S matrix elements within the theory of Ref.~\cite{Paczos:2023mof} - we believe there is no form of the FP that one can currently use~\cite{seminar}. Therefore, in the next section, we limit our discussion of tachyon FP to the DS proposal and show that it violates unitarity.

\section{Feynman Propagator}
\label{sec:Feynman_propagator}
\begin{figure}[tb]
    \includegraphics[scale=0.262]{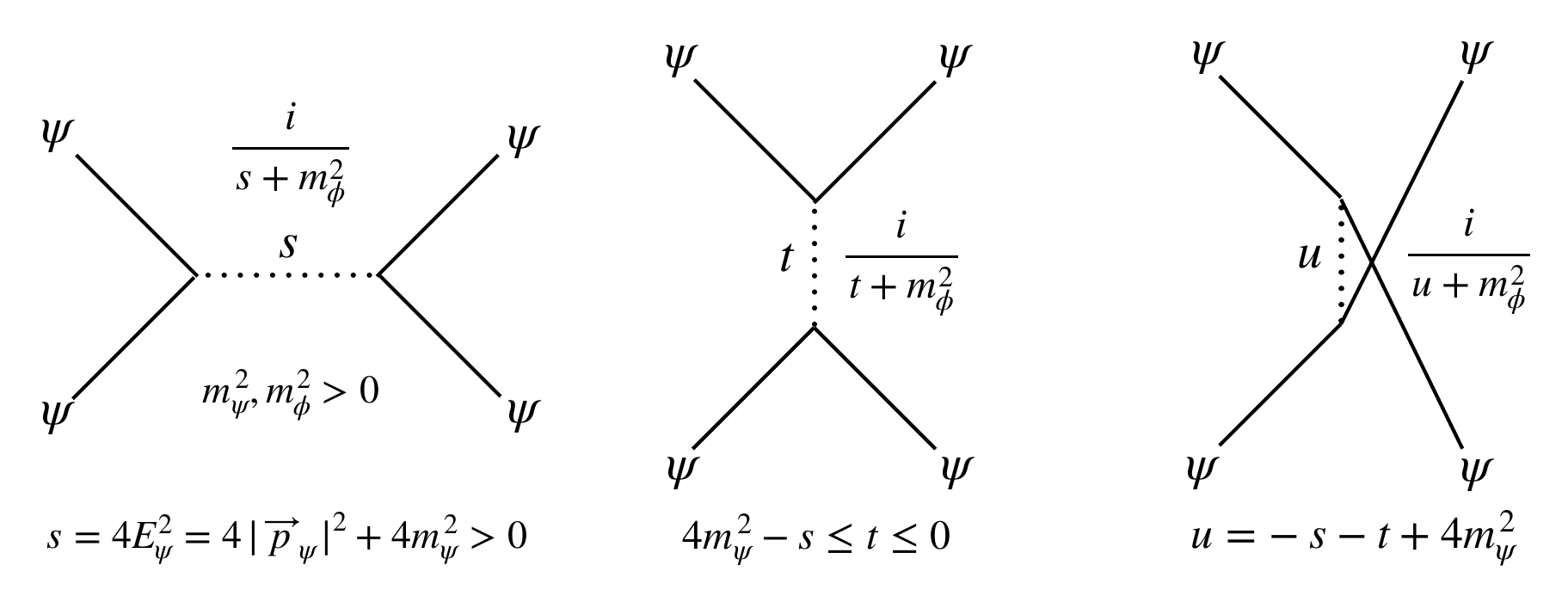}
    \caption{
        Kinematical divergences in the tree-level scattering of on-shell subluminal particles $\psi$ mediated by virtal tachyon $\phi$. 
        Poles are hit in the $t$ and $u$ channels.
        }
    \label{fig:tree_poles}
\end{figure}

\begin{figure*}[tb]
    \includegraphics[scale=0.185]{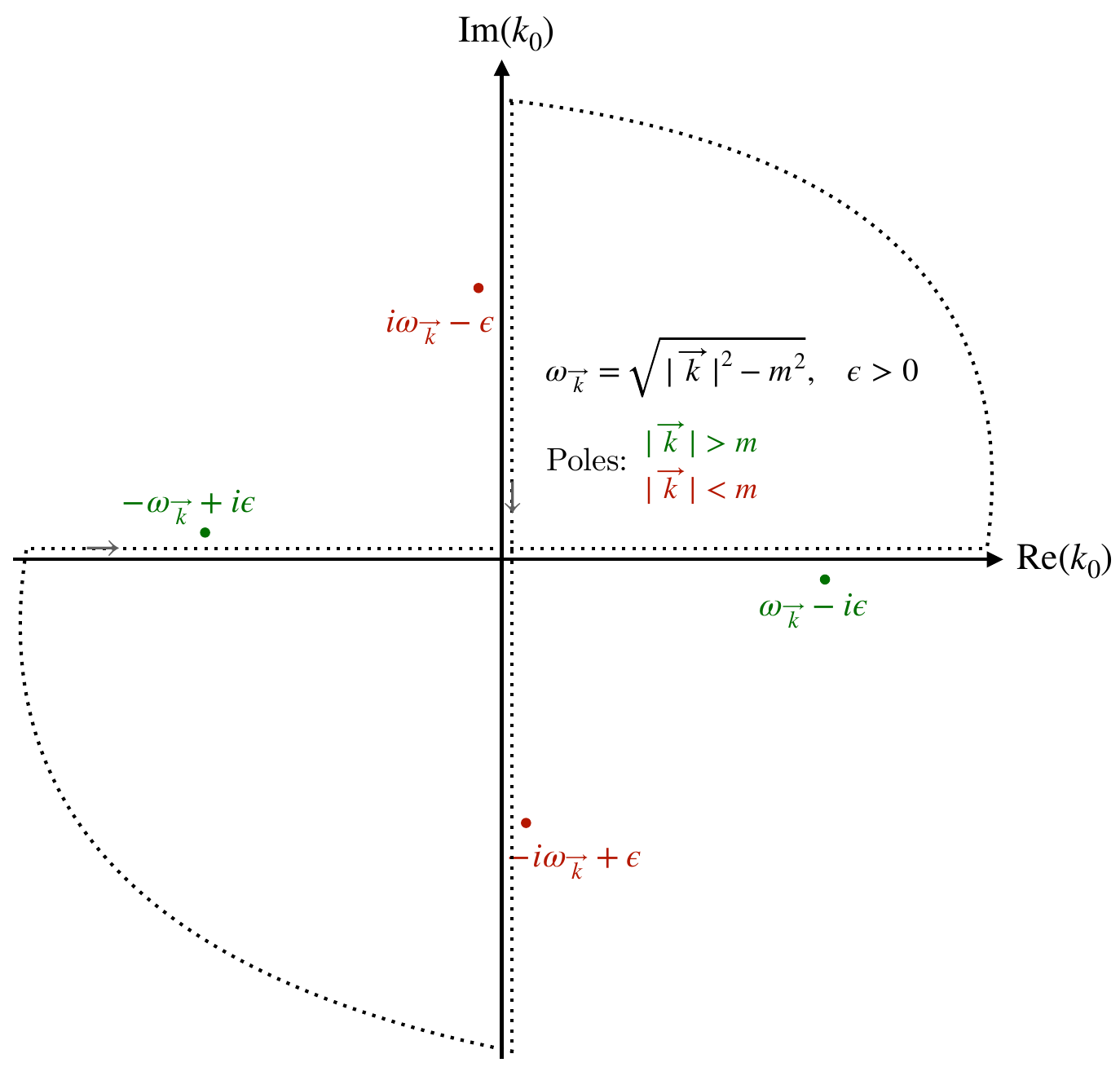}\hspace*{0.3cm}\includegraphics[scale=0.454]{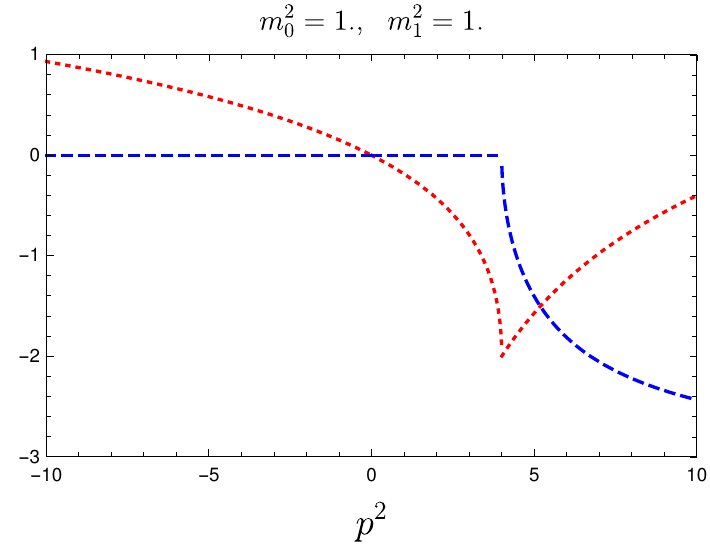}\hspace*{0.3cm}
    \includegraphics[scale=0.454]{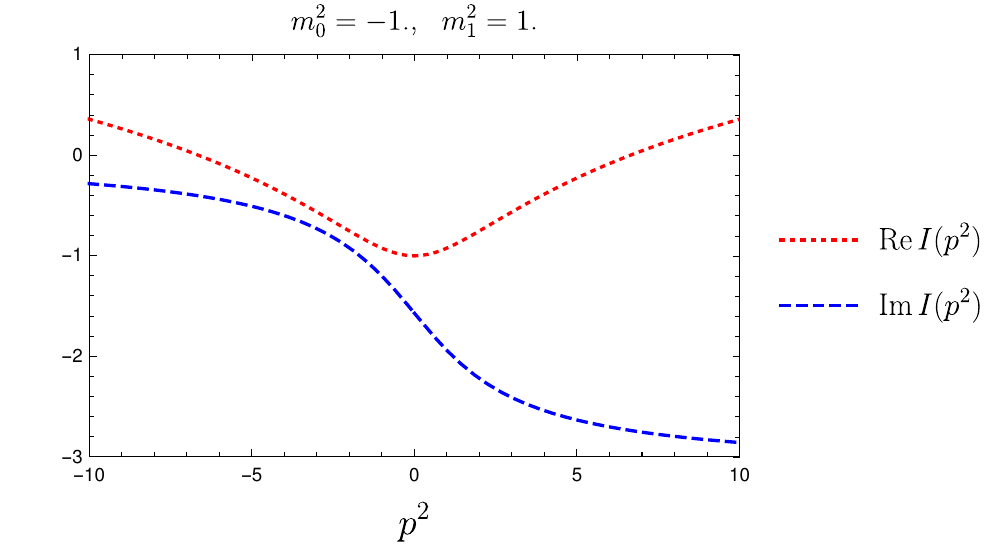}
    \caption{
        Left: Wick rotation for 1-loop computations involving virtual tachyons. 
        In the middle and right panels, we show the real and imaginary parts of the finite part of the two-point TOCF of $\psi$ for positive and negative values of $m_0^2$, respectively. 
        In the latter case, $\mathrm{Im}\,I(p^2)\neq 0$ for any finite value of $p^2$.
        The $\phi$'s propagator is given by \cref{eq:Feynman_1}.
        }
    \label{fig:Real_Imaginary}
\end{figure*}
The following equation was given in~\cite{Dhar:1968hkz} for the FP:
\be
    &\Delta_F(x-y) = \int \frac{d^4 k}{(2\pi)^4}\, \frac{i\, e^{-ik(x-y)}}{k^2+m^2+i\epsilon}\,,
\label{eq:Feynman_1}
\ee
where it was argued that the physical FP should be obtained by extending the two-point TOCF by the virtual tachyons with $|\vec{k}|\,<m$. 
No computation or proof was given, only an analogy to QED in the Coulomb gauge. We will show that i) this analogy is flawed and ii) the virtual tachyons actually do not decouple and become propagating states, violating unitarity.

In QED in the Coulomb gauge, the photon propagator is indeed not LI. This is not surprising because this gauge introduces a preferred reference frame, where the time component of the four-potential is time-independent in the vacuum. 
Moreover, the Hamiltonian contains an instantaneous Coulomb interaction. 
On the other hand, one can indeed use the covariant photon propagator, obtained as if by adding the non-physical degrees of freedom: scalar and longitudinal photons; see discussion in~\cite{Sakurai}. However, the key property that allows such modification is that the contributions to any amplitude from the scalar and longitudinal photons have the same absolute magnitude but opposite sign. This holds because, due to the gauge invariance, the QED current is conserved. As a result, the non-physical degrees of freedom can be allowed to propagate as virtual states because they decouple from any observable. Therefore, including their contributions in the propagator, which restores its LI, is mathematically equivalent to adding a zero. 
On the other hand, tachyons do \textit{not} have gauge symmetry, and nothing guarantees such decoupling. In fact, tachyons are scalars, not massless vectors - there is no redundancy in their description, as is the case of the photon, which only has two physical degrees of freedom, but is described by a four-vector. Instead, the $|\vec{k}|\,<m$ modes were excluded because they lead to exponentially increasing terms in the wave function normalization, which violates unitarity. 

We will show by direct calculation that using $\Delta_F$ defined by \cref{eq:Feynman_1}, which is indeed a LI Green's function, as FP will result in unitarity violation.
The simplest example takes place already at the tree level - it is $2\to 2$ elastic scattering shown in \cref{fig:tree_poles}.
In the center of mass frame, the subluminal scalar $\psi$, which interacts with tachyons by Yukawa-like interaction - see Eq. 22 in~\cite{Paczos:2023mof} - can transfer virtual momentum to the tachyon $\phi$ in such a way, that either $t=-m_\phi^2$ or $u=-m_\phi^2$. In fact, this will take place at infinitely many kinematical configurations provided that $|\vec{p}_\psi|\geq m_\phi/2$. Note that these divergences are physical - they generically do not lie on the edges of the allowed kinematical region ($t=0$ or $t=4m_\psi^2-s$), which correspond to the forward and backwards scatterings.
Therefore, tachyon divergences are different that the divergences in QED, where the photon propagator also can be hit. 
This is due to the fact that the latter divergences are not observable in the sense that, due to long-range nature of QED, the total cross-section is infinite, and the forward scattering is not a good observable since there is no way to actually measure it.
On the other hand, for tachyons this just shows breakdown of the DS theory, since one would have to exclude infinitely many kinematical configurations and there is no way to ``renormalize'' them since they are purely kinematical.
Actually, this property has already been noted in~\cite{Mrowczynski:1983nf}, where also volume of the phase space of tachyons has been calculated. In particular,~\cite{Mrowczynski:1983nf} found that zero-energy tachyons have infinite phase space, which also suggests that tachyon interactions can lead to instabilities.

On the other hand, one can give a more general argument based on the optical theorem.\footnote{By contraposition, if the optical theorem does not hold, the S matrix is not unitary, and we are done.} 
Let us consider the mass renormalization of a state $\psi$, which is the lightest subluminal particle in the spectrum, due to the Yukawa-like interaction with a virtual tachyon. 
For clarity, we discuss the case when $\psi$ is a scalar, while the fermion case can be analyzed in an analogous way. 
Our goal is to study the behavior of the amplitude $-i \mathcal{M}$ as a function of $p^2 = p_\psi^2$ for fixed values of parameters $m_0^2 = m_\phi^2$ and $m_1^2 = m_\psi^2>0$. The sign of $m_0^2$ is not fixed - it is positive when $\phi$ is an ordinary particle and negative when $\phi$ is a tachyon. Since our expression is UV divergent, we use dimensional regularization to isolate the finite part $I(p^2)$, which is relevant to our analysis. We combine the denominators using the Feynman-Schwinger trick, and we evaluate the integral using the Wick rotation - see the left panel of \cref{fig:Real_Imaginary} for a justification. As a result, we obtain
\begin{widetext}
    \be
        -i \mathcal{M} &= (g \Lambda)^2 \mu^{2\epsilon} \int \frac{d^D k}{(2\pi)^D} \frac{1}{(k^2-m_0^2+i\epsilon)((p+k)^2-m_1^2+i\epsilon)} = (g \Lambda)^2 \mu^{2\epsilon} \int_{0}^1 dx \int \frac{d^D l}{(2\pi)^D} \frac{1}{(l^2-\Delta)^2} \\
        &= 
        \frac{i(g \Lambda)^2}{16\pi^2} \mu^{2\epsilon} \,(4\pi)^{\epsilon} \,\Gamma(\epsilon) \int_{0}^1 dx \frac{1}{\Delta^\epsilon} = \frac{i(g \Lambda)^2}{16\pi^2} \left[ \Big(\frac 1\epsilon -\gamma_E + \ln\left(4\pi \mu^2\right) \Big) - \int_{0}^1 dx \log\left( -x(1-x)p^2+(1-x)m_0^2+xm_1^2-i\epsilon \right) \right] \\
        &= \frac{i(g \Lambda)^2}{16\pi^2} \left[\mathrm{UV} - I(p^2)\right] = \frac{i(g \Lambda)^2}{16\pi^2} \left[ \mathrm{UV} - \int_{0}^1 dx \Big( \log\left(p^2-i\epsilon\right) + \log\left((x-x_{+})(x-x_{-}) \right) \Big) \right] \\
        &= \frac{i(g \Lambda)^2}{16\pi^2} \left[\mathrm{UV} - \Big( \log\left(p^2-i\epsilon\right) + (1-x_{+})\log\left(1-x_{+}\right) + (1-x_{-})\log\left(1-x_{-}\right) -2 + x_{+}\log(-x_{+}) + x_{-}\log(-x_{-}) \Big)\right] \,,
    \label{eq:Wick_1}
    \ee
\end{widetext}
where $D=4-2\epsilon<4$, $l=k+px$, $\Delta(x)=-x(1-x)p^2+(1-x)m_0^2+xm_1^2-i\epsilon$, and $\Delta(x_\pm)=0$ for $x_{\pm} = \frac12 \left(1+\frac{m_0^2-m_1^2}{p^2}\right) \pm \sqrt{\left(1+\frac{m_0^2-m_1^2}{p^2}\right)^2 - 4 \frac{m_0^2-i\epsilon}{p^2}}$. 
We kept the leading terms, $\Gamma(\epsilon) \simeq \frac 1\epsilon -\gamma_E$ and $A^\epsilon \simeq 1+\epsilon \,\log(A)$. 

To cross-check \cref{eq:Wick_1}, we also calculated the imaginary part of $\mathcal{M}$ directly. 
It is given by the following simple integral, which, \eg, Mathematica can evaluate:
\be
    \mathrm{Im}(\mathcal{M}) = -\pi\int_0^1 dx \, \theta\left[x(1-x)p^2-(1-x)m_0^2-xm_1^2 \right].
\label{eq:Wick_2}
\ee
To obtain it from \cref{eq:Wick_1}, we used $\ln(-A-i\epsilon)=\ln(A)-i\pi$ for positive $A$ and positive infinitesimal $\epsilon$.

In the middle and right panels of \cref{fig:Real_Imaginary}, we show our results. The region relevant to the mass renormalization is $0<p^2<4 m_\psi^2$. 
Indeed, when $m_\phi^2>0$, the imaginary part of $I(p^2)$ vanishes in this region, making $\mathcal{M}$ purely real, as required by the optical theorem for a stable $\psi$.
On the other hand, for $m_0^2<0$, the imaginary part of $I(p^2)$ vanishes only in the $p^2\to -\infty$ limit, which means that the propagator of $\psi$ will develop a finite imaginary part. 
This means that $\psi$ is actually an unstable particle, contrary to our initial assumption. 
This contradiction means that the theory is mathematically inconsistent or interactions of virtual tachyons described by \cref{eq:Feynman_1} make any subluminal particle they interact with unstable.
Actually, if tachyon have self-interactions, in this case cubic, we can repeat this procedure for tachyon field itself - the only difference is to put $m_1^2=m_0^2<0$ in \cref{eq:Wick_1,eq:Wick_2}.
Again, we find that for any finite value of $p^2$ tachyon acquires a finite decay width, hence the tachyon self-interactions turn a tree-level stable tachyon into an unstable state (assuming the FP is given by \cref{eq:Feynman_1}).

What is more, since energy conservation prohibits decay of $\psi$ at rest into itself and a positive energy tachyon, such decays would have to involve the $|\vec{k}|\,<m$ degrees of freedom, which  violate unitarity. 
While we performed explicit calculations for the Yukawa-like interactions, we expect that a similar argument would apply to any interactions of tachyons, hence only QFT of free tachyons in the DS theory could be unitary. 
However, since the status of the FP is unclear in Ref.~\cite{Paczos:2023mof}, the results shown in this section may not apply to that work.

On the other hand, if the DS propagator is adapted, than the process considered in Ref.~\cite{Paczos:2024mfx} violates unitarity, since the thresholds of the tree-level and loop-level decay widths are different (in addition to the fact that, by the arguments given in \cref{sec:twin_space}, the QFT of tachyons considered in~\cite{Paczos:2024mfx,Paczos:2023mof} is not quantum). 
Therefore, without specifying the FP for tachyon, the setup proposed in~\cite{Paczos:2024mfx} cannot be a Higgs mechanism toy model. For completeness, we note that there are typos in Eq. 13-16 therein, which, however, do not change the final result given by Eq. 12, since correct expression for the amplitude was used (the term furthest to the right in Eq. 16).

\section{LSZ formalism for tachyons}
\label{app:S_matrix}
The question, which we did not address so far, is whether a satisfactory scattering theory of tachyons can be formulated at all.\footnote{In this section, we consider tachyon scattering theory as a model-independent. We use the positive-energy mass-shell measure, $d^3\vec{k}/(2\omega_{\vec{k}})=d^3\vec{k}/(2\sqrt{|\vec{k}|^2-m^2})$, in a fixed frame.}

While detailed analysis is beyond our scope, we examine the textbook discussion of the LSZ formalism~\cite{Coleman:2011xi,Srednicki:2007qs}.
We show that restricting the modes to $|\vec{k}|>m$ and making analogous assumptions as for the $m^2>0$ case, is not sufficient to establish the LSZ reduction formula.

A key step is showing that the field becomes free (in the sense of weak limit) at past/future infinity.
As one of the steps, Coleman applies the Riemann--Lebesgue lemma to the multiparticle term $R_\psi(t)$ defined as
\be
    R_\psi(t) &= \int_{\rm multi}d\nu(n)\, \frac{C_\psi(n)}{\omega_{\vec P_n}} e^{it(E_n-\omega_{\vec P_n})} 
\,,
\ee
where $d\nu(n)$ denotes the $n$-particle measure, $\phi_{\rm int}$ is the interacting interpolating field, and $C_\psi(n) = \frac12\langle\psi|n\rangle\, \tilde f(\vec P_n)(E_n+\omega_{\vec P_n}) \langle n|\phi_{\rm int}(0)|0\rangle$.
A sufficient integrability condition is that $C_\psi\in L_1(d\nu)$, which allows to apply the Riemann--Lebesgue lemma, showing the multiparticle term vanishes asymptotically,
\be
    \lim_{t\to\pm\infty} R_\psi(t) &= 0
\,.
\ee
On the other hand, for tachyons, the corresponding sufficient absolute-integrability condition is
\be
    \int d\nu(n)\, \frac{|C_\psi(n)|}{\omega_{\vec P_n}}<\infty
\,.
\ee
This is stronger local integrability condition because $1/\omega_{\vec P_n}$ is unbounded. 

A second, independent issue is that for tachyons, the kinematic separation inequality, $E_{\rm multi}(\vec P)>\omega_{\vec P}$, may not hold even when every constituent has momentum greater than the mass. 
For example, for $\lambda/4! \phi^4$ coupling, an initial state with momentum
\be
    p=(E,0,0,K)\,,\qquad K^2=E^2+m^2,\qquad E>0\,,
\ee
can decay into $3$ other tachyons; for $j=0,1,2$ define
\be
    q_j= \left( \frac E3,\, \frac{2\sqrt2m}{3}\cos\frac{2\pi j}{3},\, \frac{2\sqrt2m}{3}\sin\frac{2\pi j}{3},\, \frac K3 \right)
\,.
\ee
These satisfy
\be
    &\sum_jq_j=p\,,
    \qquad
    q_j^2=-m^2\,,
    \qquad
    \\
    &q_j^0=\frac E3>0\,,
    \qquad
    |\vec q_j|^2=m^2+\frac{E^2}{9}>m^2
\,.
\ee
Thus, the one-tachyon dispersion lies within a three-tachyon continuum, and the decay into three equal-mass tachyons is kinematically allowed at every positive initial energy. 
If the corresponding decay amplitude is nonzero on this phase space, the tachyon is unstable and cannot serve as a standard LSZ asymptotic state. 

In summary, we have shown that one needs to impose stronger local integrability requirements and that the existence of exact stable one-particle states requires a separate dynamical justification, because existence of stable one-particle asymptotic states is no longer protected by kinematics.

\section{Conclusions}
\label{sec:conclusions}
Tachyons are usually considered undesirable artifacts that signal inconsistency or instability of the theory. While recent works on tachyons~\cite{Dragan:2019grn,Dragan:2022txt} found an interesting connection between superluminal observers and the foundations of QM, which revived the topic and made tachyons into a well-motivated subject of theoretical investigations, our results agree with such a sentiment. 

In particular, we have shown that the theory constructed in Ref.~\cite{Paczos:2023mof} is actually a classical theory with no quantum dynamics in the twin space, the DS FP violates unitarity, and the LSZ asymptotic condition in any covariant tachyon QFT cannot be proved just by replacing non-normalizable plane waves with normalizable wave packets.
These results suggest that constructing a covariant QFT of tachyon, defined as a scalar stable particle with negative mass squared, seems impossible in Minkowski space without giving up unitarity or other fundamental QFT principle.

\acknowledgments
This work was supported by the Institute for Basic Science under the project code, IBS-R018-D1. 

We thank Sławomir Breiter and Jan Dereziński for the opportunity to give seminars on this topic at Adam Mickiewicz University, Poznań and University of Warsaw~\cite{seminar_2}, respectively.
We thank Ryszard Horodecki for correspondence regarding Ref.~\cite{Horodecki:2023hvf}, useful comments, and for the opportunity to give a seminar on this topic at University of Gdańsk.
We also thank Jan Dereziński for useful discussions and Stanisław Mrówczyński for correspondence regarding Ref.~\cite{Mrowczynski:1983nf} and useful remarks.
Finally, we thank the anonymous Referee for useful comments, and the Authors of Ref.~\cite{Paczos:2023mof}, especially Andrzej Dragan, for criticism and discussions, which greatly improved our understanding of the problem. 

\bibliography{bibliography}

\end{document}